\documentclass[12pt,onecolumn]{article}
\usepackage[dvips]{graphicx,color}
\newcommand{\numu}{\ensuremath{\nu_\mu}}
\newcommand{\nue}{\ensuremath{\nu_\mathrm{e}}}

\newcommand{\numubar}{\ensuremath{\bar \nu_\mu}}
\newcommand{\nuxcarbon}{\ensuremath{\nu_\mathrm{x} \hspace{0.05in} ^{12}\mbox{C} \rightarrow \nu_\mathrm{x} \hspace{0.05in} ^{12}\mbox{C}^{*}(15.11)}}
\newcommand{\numucarbon}{\ensuremath{\nu_{\mu} \hspace{0.05in} ^{12}\mbox{C} \rightarrow \nu_{\mu} \hspace{0.05in} ^{12}\mbox{C}^{*}(15.11)}}
\newcommand{\nuecarbon}{\ensuremath{\nu_{\mathrm{e}} \hspace{0.05in} ^{12}\mbox{C} \rightarrow \mbox{e}^- \hspace{0.05in} ^{12}\mbox{N}_{\mathrm{gs}}}}
\newcommand{\ev}{\ensuremath{\mbox{eV}^2}}
\newcommand{\dm}{\ensuremath{\Delta \mbox{m} ^2}}
\newcommand{\st}{\ensuremath{\sin^2 2 \theta}}
\newcommand{\fc}{\ensuremath{\phi_{\nu_{\mu}} \times \sigma_{\mathrm{NC}}}}

\newcommand{\mus}{\ensuremath{\nu_{\mu} \rightarrow \nu_\mathrm{s}}}

\begin {document}

\title{\bf Measuring Active-Sterile Neutrino Oscillations with a Stopped Pion Ne
utrino Source}

\author{G.T.\ Garvey, A.\ Green, C.\ Green, W.C.\ Louis, G.B.\ Mills, \\
G.\ McGregor, H.\ Ray, R.\ Schirato,  R.G.\ Van de Water, \\
D.H.\ White\\
Los Alamos National Laboratory \\
LA-UR-04-8716}

\date{\today}

\maketitle

\abstract
The possible existence of light sterile neutrinos is of great
interest in many areas of particle physics, astrophysics, and
cosmology.  Furthermore, should the MiniBooNE experiment at Fermilab
confirm the LSND oscillation signal, then new measurements are
required to identify the mechanism responsible for these
oscillations. Possibilities include sterile neutrinos, CP or CPT
violation, variable mass neutrinos, Lorentz violation, and extra dimensions.  In this
paper, we consider an experiment at a stopped pion neutrino source to
determine if active-sterile neutrino oscillations with $\Delta \mbox{m}^2$ greater
than 0.1 \ev can account for the signal. By exploiting stopped
$\pi^+$ decay to produce a monoenergetic
$\nu_\mu$ source, and measuring the rate of the neutral current reaction 
\nuxcarbon\ as a function of distance
from the source, we show that a convincing test for active-sterile
neutrino oscillations can be performed.

\section{Introduction}

The understanding of neutrino mass has undergone a revolution over the
last ten years. Neutrino mass, via oscillations, is now experimentally
established \cite{mass}.  The Standard Model may now need to
accommodate right handed neutrinos, $\nu_{R}$, that are sterile with
respect to the weak interaction.  The new issue to be addressed concerns
the mass spectrum of these sterile neutrinos.  Although many theorists
postulate that $\nu_{R}$ are heavy via the see-saw mechanism
\cite{Gmann}, there is no experimental evidence for the mass spectrum
of sterile neutrinos or for the number of generations \cite{noseesaw}.
In fact, light sterile neutrinos have been invoked to explain many
puzzles, such as the LSND result \cite{(3+2)}, the solar neutrino
energy spectrum anomaly \cite{solar}, the R-process in type II
supernovae \cite{supernova}, pulsar kicks \cite{pulsar}, dark matter
\cite{dmatter}, the perplexing dark energy \cite{denergy}, and most
recently, extra dimensions \cite{extraD}.  Thus, much experimental and
theoretical justification exists to motivate a search for light
sterile neutrinos above the mass of $0.1$\,\ev.

The LSND result points to active-sterile neutrino oscillations in the
\dm\ range of 0.1 to 2\,\ev\ \cite{lsnd}. A recent analysis of a (3+2)
sterile neutrino model indicates the possibility of a second sterile
neutrino solution around 20\,\ev\
\cite{(3+2)}.  In addition, a supernovae R-process analysis favors a 
$3 < \dm < 70$\,\ev\ 
sterile neutrino \cite{supernova}.  We propose an experiment that is
capable of the direct detection of active-sterile neutrino
oscillations in this mass region.  If we interpret 
the LSND result as an intermediate sterile state, i.e.\
$\bar{\nu}_{\mu}
\rightarrow \bar{\nu}_\mathrm{s} \rightarrow \bar{\nu}_\mathrm{e}$, where 
the observed
LSND oscillation probability is $P(\bar{\nu}_{\mu}
\rightarrow
\bar{\nu}_\mathrm{e}) = (0.264 \pm 0.081) $\% \cite{lsnd}, then  coupling this with 
short baseline reactor
experimental limits on $\nu_\mathrm{e}$ disappearance of $P(\bar{\nu}_{\mathrm{e}} \rightarrow \bar{\nu}_\mathrm{s}) < 10$\% \cite{reactor},
we can predict the $\nu_{\mu}$ disappearance to sterile
neutrinos to be $P(\bar{\nu}_{\mu} \rightarrow \bar{\nu}_\mathrm{s}) > 10$\%.
The above arguments set the scale for the \dm\ and \st\ sensitivity
that needs to be achieved.

\section{Beyond the Standard Model}

Given the prospect of a new intense stopped pion neutrino source,
together with a positive signal from MiniBooNE \cite{miniboone}, a next 
generation
short baseline neutrino experiment needs to be carefully considered.
A neutrino detector at the Spallation Neutron Source (SNS) has
already been discussed
\cite{VanDalen} that  demonstrates how an improved measurement in the
LSND/MiniBooNE region can be made.  Improving the
\dm\ and \st\ measurement of any positive signal from MiniBooNE would be 
important.  However, the nagging question would still remain as to
the new physics behind the three distinct \dm\ regions that have been
measured with solar, reactor, atmospheric, and short baseline detectors.
Several possibilities at present exist, such as sterile neutrinos
\cite{(3+2)}, CP or CPT violation \cite{cpt}, mass varying neutrinos
\cite{denergy}, Lorentz violation \cite{lorentz}, and extra dimensions \cite{extraD}.

The CP or CPT violation possibility could be readily tested with a
MiniBooNE/LSND style detector coupled with a short duty-factor 
stopped pion beam which allows,
via timing cuts, a separation of \numu\ and \numubar\ events.  Thus, a
direct comparison of oscillation probabilities from these two
simultaneous data sets will allow a test of CPT and CP violation
models.  Mass varying neutrinos can be tested by changing the density
of material between the source and detector.  The Lorentz violation 
model can be tested by looking for sidereal variations in the 
neutrino flux.  These three tests will be discussed in more detail
in a future paper.

This paper discusses the first possibility, that of sterile neutrinos.
To demonstrate active-sterile neutrino oscillations directly, and
achieve the best sensitivity, requires a two detector setup and a
stopped pion neutrino source.  The following sections describe how
this can be done.

\section{Intense Stopped Pion Neutrino Sources}

To detect sterile neutrino effects in the mass range of LSND and
above requires an intense source of well characterized neutrinos. 
Stopped pion decay from a low energy proton beam is such a source.
Neutrinos from stopped pion decay
have a well defined flux, well defined energy spectrum, and low backgrounds.
Fortunately, accelerators that can provide such a source 
are either being built or
are being proposed.  The first such accelerator is the 1.4\,MW, 1.3\,GeV,
short duty-factor Spallation Neutron Source (SNS) \cite{ornl},
which is currently being built and will be fully commissioned by 2008.
The second is a proposal for a 2\,MW, 8\,GeV, Proton Driver (PD) at FNAL
\cite{fnal}.

The dominant decay scheme that produces neutrinos from a stopped pion
source is 
\begin{equation}
 \pi^+ \rightarrow \mu^{+} \numu, \hspace{0.5cm} \tau = 26 \hspace{0.1cm} \mbox{nsec}
\end{equation}
followed by
\begin{equation}
\mu^+ \rightarrow e^+ \numubar \nu_e, \hspace{0.5cm} \tau = 2.2 \hspace{0.1cm} \mu \mbox{sec}.
\end{equation} 

\begin{figure}
\centering
\includegraphics[scale=0.75]{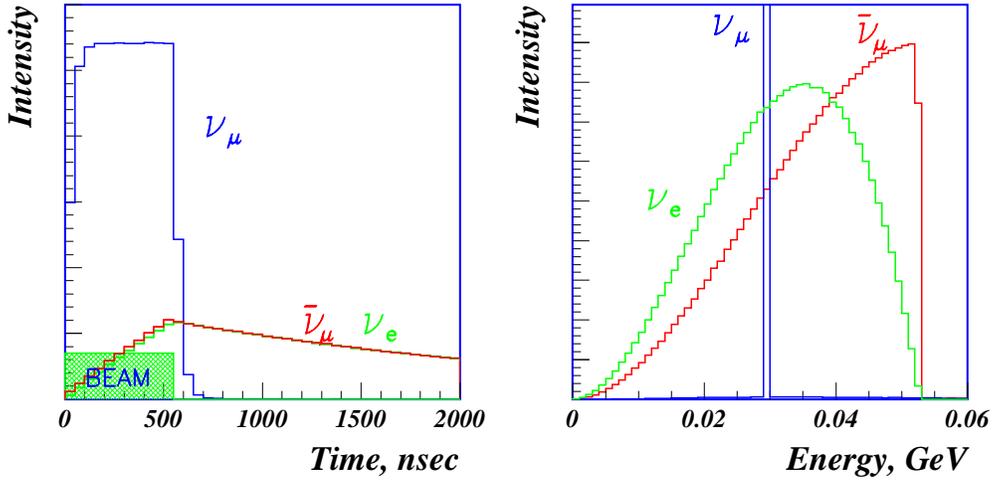}
\parbox{6in}{
\caption{ \small The neutrino time and energy spectra 
of the different neutrino species produced isotropically
from a stopped pion source \cite{ornl}.}
\label{sns_flux} }
\end{figure}

The neutrinos from stopped $\pi^-$'s are highly suppressed because the
negative pions are absorbed in the surrounding
material.  Thus, neutrinos from the $\pi^-$ decay chain are
significantly depleted and can be estimated from the measured \numu,
\numubar, and \nue\ flux. Figure \ref{sns_flux} shows the neutrino
time and energy spectra from a stopped pion source.  As shown in the
right hand plot, the \numu\ energy is monoenergetic ($E_{\numu} =
29.8$\,MeV), while the
\numubar\ and \nue\ have known Michel decay distributions with an end point energy of
52.8\,MeV.  Furthermore, the left hand plot shows the SNS beam timing
and the time distributions for the three neutrino species.  One of the
advantages of the SNS relative to the LAMPF accelerator is the
three orders of magnitude shorter beam time of 695\,nsec (the FNAL PD
will have a similar beam time).  With a simple beam-on timing cut, 
one can obtain a fairly pure
\numu\ sample, with only a 14\% contamination of \numubar\ and \nue\ each.  
This remaining background is easily measured and subtracted.  

Table \ref{table1} shows the expected proton rates for both SNS and
FNAL beamlines, normalized to a full year of running,
i.e.\ $3.15\times10^{7}$ seconds.  The FNAL Proton Driver proposal is
broken down into 8\,GeV, 2\,MW and 5\,GeV, 1.25\,MW beamlines.  The FNAL
8\,GeV option will provide about 2.5 times more protons per year than
the SNS.  However, the FNAL PD is only a proposal, while the SNS
is under construction and will be operational by 2008.  This makes the SNS 
a more
timely option.  Furthermore, the SNS is planning for 2014 an upgrade
which will deliver 3\,MW to two sources, making the interesting
situation of multiple baselines with a single detector.

A key component of the sterile neutrino measurement
is the physical size  of the stopped pion source, which adds an uncertainty
to the neutrino path length. For the SNS, the compact liquid mercury
target will contribute approximately 25\,cm (FWHM) to the neutrino path
length uncertainty.  The FNAL source size should be of similar
dimensions to minimize neutrino path length uncertainties.

\begin{table}[t]
\centering
\vspace{5mm}
\begin{tabular}{|c|c|c|c|}
\hline
 & FNAL (8 GeV) & FNAL (5 GeV) & SNS \\ \hline
 P/yr         & $1.6\times 10^{22}$ & $1.6\times 10^{22}$ & $6.7\times 10^{22}$ \\ \hline 
 DAR $\nu (\nu/P)$ & 1.5  & 0.9  & 0.13 \\
 DAR $\nu (\nu/yr)$ & $7.3\times10^{22}$ & $4.4\times 10^{22}$ & $2.9\times 10^{22}$ \\ \hline 
\end{tabular}
\parbox{6in}{
\caption{\small Proton intensities at FNAL and SNS. The numbers are taken from \cite{fnal}, assuming $3.15\times10^{7}$ s/yr operation. }
\label{table1}}
\end{table}

\section{Active-Sterile Neutrino Oscillation Measurements}

The direct observation of active-sterile neutrino oscillations with 
$\dm > 0.1$\,\ev\
can be achieved because of two key features of a short duty-factor stopped pion
neutrino source.  Firstly, all active neutrino species ($\mbox{x} = \nue, \numu, \nu_{\tau}$)
can be efficiently measured using the superallowed neutral current reaction \nuxcarbon\
\cite{xsecarbon}, where the $\mbox{C}^{*}(15.11)$ state has a $92 \pm 2$\%  branching
fraction to the ground state \cite{carbondecay}. A deficit of the
neutral current rate in $^{12}\mbox{C}$ from expectation can only be from
oscillations to sterile neutrinos.  Secondly, using beam timing cuts,
we can extract a high purity sample of
29.8\,MeV monoenergetic \numu\ events. In (3+1) models with 
$\dm > 0.1$\,\ev, active-sterile oscillations can be
approximated by two neutrino mixing. The oscillation probability can
be expressed as:
\begin{equation}
P( \numu \rightarrow \numu ) = 1 - \st \sin^2 ( 1.27 \hspace{0.1cm} \dm 
\frac{\mbox{L}}{29.8} ) .
\end{equation} 

\begin{figure}
\centering
\includegraphics[scale=0.60]{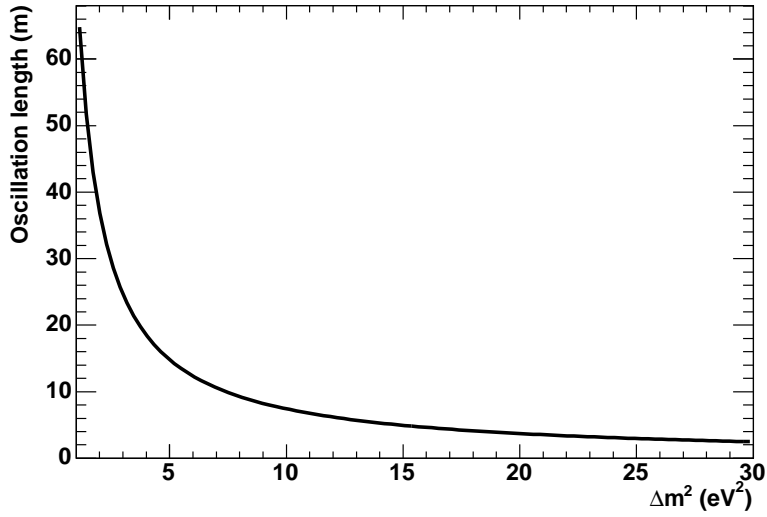}
\parbox{6in}{
\caption{\small Oscillation length as a function \dm\ where $E_{\nu_\mu}= 29.8$\,MeV.}
\label{f1}}
\end{figure}

\begin{figure}
\centering
\includegraphics[scale=0.60]{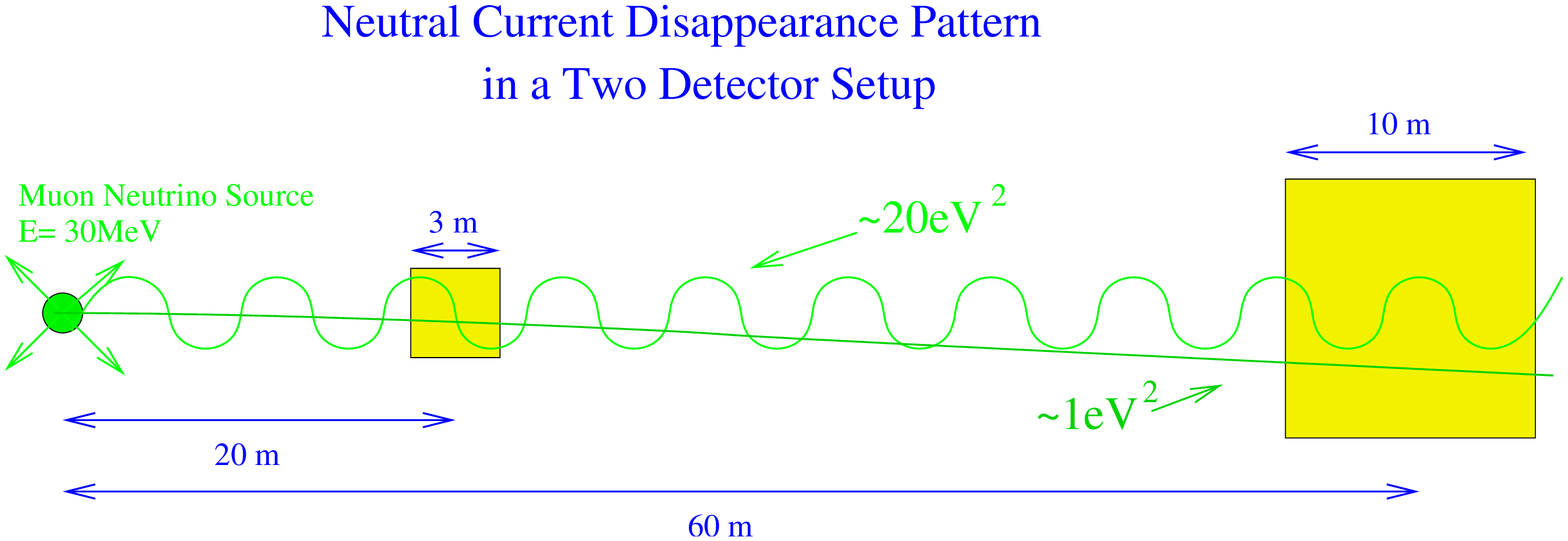}
\parbox{6in}{
\caption{\small Neutral Current \nuxcarbon\ disappearance patterns for a two 
detector setup.}
\label{f2}}
\end{figure}

Given these two features, we can observe active-sterile neutrino oscillations
by measuring changes in the \nuxcarbon\ rate as a function of L
(in meters).  Figure \ref{f1} shows the oscillation length as a
function of \dm.  Thus, a two detector configuration of the 
appropriate size and distance can achieve the desired
sensitivity.  A near detector greater than three meters in length and about
10-20 meters from the source will be sensitive to $ \dm > 20$\,\ev.  A
second large detector (required to increase statistics because of decreasing 
$1/\mbox{r}^2$ neutrino flux)
at 50-100 meters will be sensitive to $ \dm > 0.1$\,\ev.  Figure
\ref{f2} shows a simple schematic of a detector setup that would allow
sensitivity to sterile neutrino oscillations using the \nuxcarbon\
channel.  
 
A possible detector configuration at the SNS would be first a 25\,ton
fiducial mass (3.5\,m on a side) liquid scintillator near detector at
18\,m distance from the neutrino source.  This is the homogeneous
detector proposed by the $\nu$-SNS collaboration to measure neutrino
carbon cross sections \cite{ornl}.  The limited fiducial mass size of
the near detector might be improved with future developments in flat
PMT technology. Next would be a 500\,ton fiducial mass liquid
scintillator (6\,m radius) far detector at 60\,m distance from the
neutrino source.  Both designs would be similar to an LSND style
detector, i.e. a mineral oil detector doped with butyl PBD
scintillator to increase light output, instrumented with 25\%
photocathode coverage.  This gives approximately a 7\% energy resolution
for reconstruction of the 15.11\,MeV gamma ray.  Both detector masses
are quoted in fiducial mass, which is the useful mass available for
reliable event reconstruction.  To estimate NC neutrino rates in the
near and far detector, we scale to the detection rate calculations in
\cite{VanDalen}, which have assumed a \numucarbon\ cross section of
$2.7 \times 10^{-42} \hspace{0.1cm}\, \mbox{cm}^2$ \cite{xsecarbon}.
The Karmen collaboration has measured this cross section with limited
statistics, getting a value of $(3.2 \pm 0.5_{\mathrm{stat}} \pm
0.4_{\mathrm{syst}}) \times 10^{-42} \hspace{0.1cm} \, \mbox{cm}^2$
\cite{karmen}, consistent with the calculated cross section.  

For a proposed experiment at FNAL, we would not have space
restrictions, allowing for larger near and far detectors.  We assume a
longer baseline, out to 100\,m, and a detector mass of 2\,ktons (8\,m
radius), which would be required to keep roughly the same statistics.
This would give a fiducial mass of 1.3\,ktons (7\,m radius).

Table \ref{tab2} shows the number of events expected per year for
the \nuxcarbon\ reaction for the two possible
configurations, one at SNS with 1.4\,MW, and the other at FNAL with 2\,MW.  The event
reconstruction efficiency was assumed to be 100\%, as the 15.11\,MeV gamma ray 
is relatively easy to detect.

\begin{table}[t]
\begin{tabular}{|c|c|c|c|c|}
\hline
Detector & Source Dist.\ (m) & FD Size (tons) & FD Length (m) & $\nu_{\mu} \hspace{0.05in} ^{12}C \rightarrow \nu_{\mu} \hspace{0.05in} ^{12}C^{*}$
events/yr \\ \hline
SNS Near      & 18   & 25 & 3 &   2056 \\
SNS Far       & 60    & 500 & 10  & 3701\\ \hline
FNAL Near     & 10   & 116 & 5 & 77806 \\
FNAL Far      & 100 & 1300 & 12 & 8720 \\ \hline
\end{tabular}
\caption{
\small Estimated \numucarbon\ events per year at the SNS (1.4\,MW) and FNAL (2\,MW) sources, assuming 100\% event reconstruction efficiency. }
\label{tab2}
\end{table}

A Monte Carlo calculation was performed to estimate the \nuxcarbon\
oscillation sensitivity, which includes smearing of 25\,cm (FWHM) from
the finite neutrino source size and 50\,cm (FWHM) from the
reconstructed position resolution for the 15.11\,MeV gamma ray.  Also
included is background subtraction of charged current (CC) and neutral
current (NC) events from \numubar\ and \nue\ in the beam window
\cite{VanDalen}.  Because we can isolate a clean sample of these
backgrounds in the time window after the beam, we will have an
excellent estimation of their number and character, e.g.\ for the SNS
far detector this is about 1300 events per year.  This background
increases the statistical error on the measured \numucarbon\ rate by
at most 45\%.  The background contribution from cosmic rays is
negligible because of the short duty factor of the beam.

\begin{figure}
\centering
\includegraphics[scale=0.55]{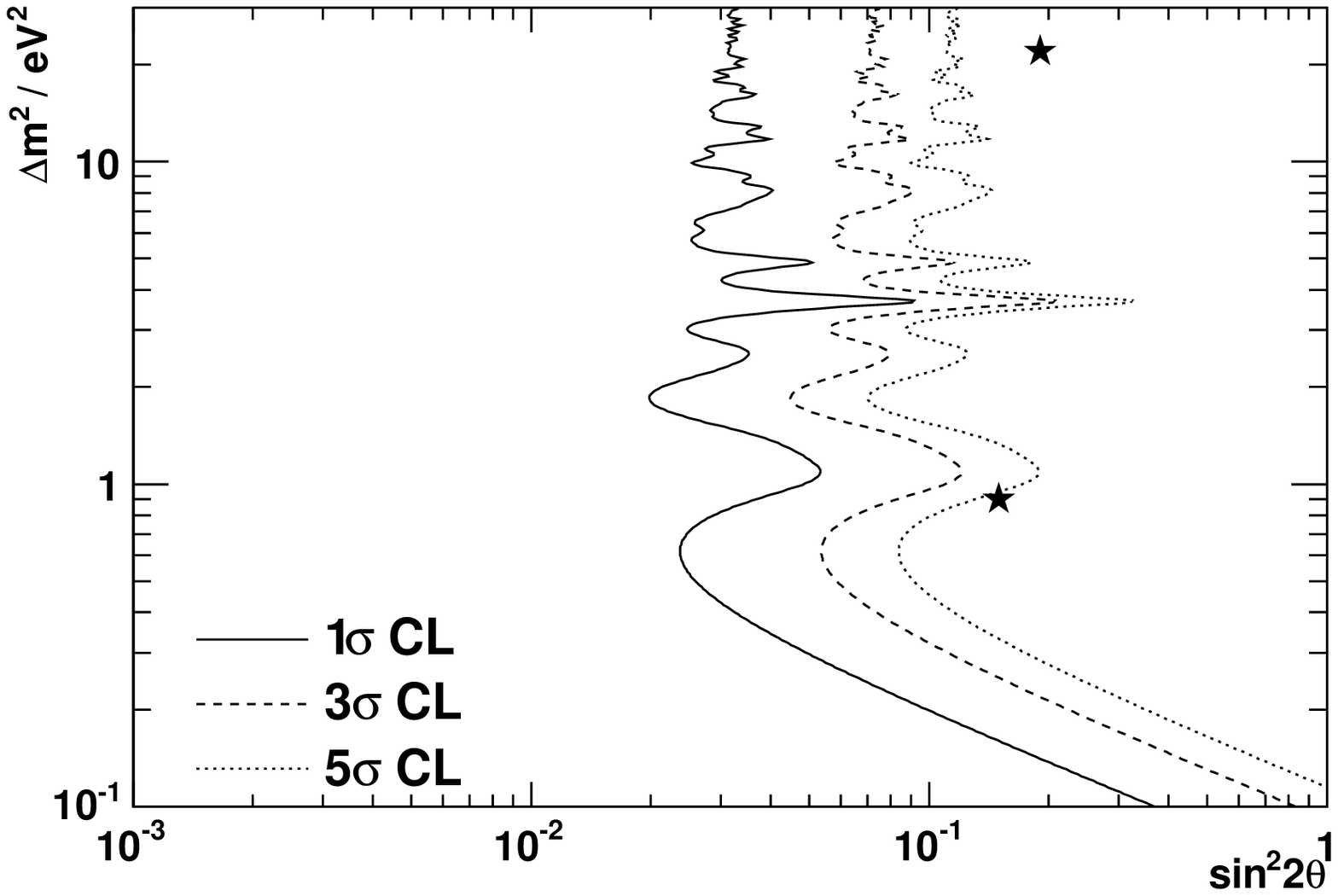}
\parbox{6in}{
\caption{\small Active-Sterile neutrino oscillation sensitivity with the NC reaction \nuxcarbon\
for SNS with two detectors, 5\% \fc\ systematic errors, and three years of running.  The two black stars correspond to the (3+2)
solutions (see text).}  
\label{f5}}
\end{figure}

\begin{figure}
\centering
\includegraphics[scale=0.55]{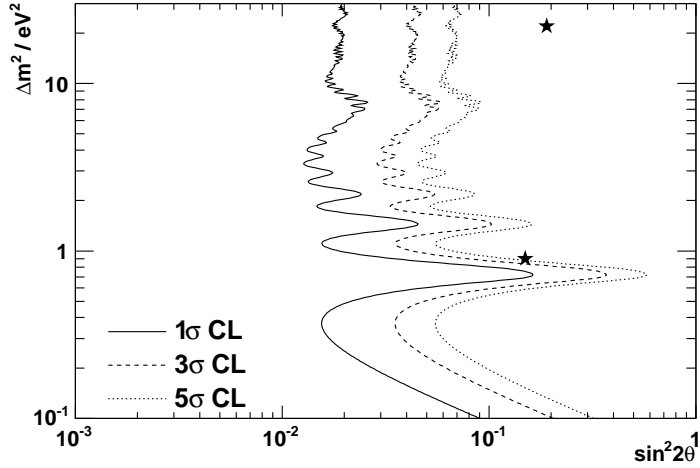}
\parbox{6in}{
\caption{\small Active-Sterile neutrino oscillation sensitivity with the NC reaction \nuxcarbon\ for FNAL PD with two detectors, 5\% \fc\ systematic errors, and three years of running.  The two black stars correspond to the (3+2)
solutions (see text).}  
\label{f6}}
\end{figure}

Figures \ref{f5} and \ref{f6} show the (3+1) active-sterile neutrino
oscillation sensitivity for a three year run at the SNS and FNAL
proton driver with two detectors and a 5\% normalization systematic errors.  
Both setups achieve a \mus\
oscillation sensitivity of $3 \sigma$ for $\dm > 0.4$\,\ev\ and $\st >
0.05$.  This sensitivity is desired if the LSND signal is due to active-sterile
neutrino oscillations.
 
The oscillation sensitivity plots are generated from a simultaneous fit
to both the L shape and to an overall event normalization.  The L
measurement systematic errors are discussed above.  The event
normalization requires knowledge of the \numu\ flux times neutral
current cross section (\fc).  From rate calculations alone, we
expect systematic errors on the order of 10\%, which is what LSND was
able to achieve \cite{lsnd}.  
Another technique is to use the charged current
\nuecarbon\ reaction to normalize the NC rate, as the final state 
A$=12$ nuclear 
form factors are isotopic analogues, leading to systematic errors in the 
NC/CC ratio of $\sim$ 3\% \cite{karmen}.  
Here we assume there are no oscillations of
\nue.  The NC/CC ratio is dominated by systematic errors since the
event rates for both reactions are large.  Besides the theoretical
ratio error (discussed above), we need to know the relative fiducial
volume efficiency and the relative detection efficiency of the $15.11$
MeV $\gamma$ to that of the CC electron plus $^{12}$N beta decay.
Based on experiences at LSND, these two errors are conservatively
estimated at 3\% each.  
This gives a total systematic uncertainly in
the NC/CC ratio of 5\%, which translates into a \numucarbon\ event
rate \fc\ systematic error of 5\%.  

\begin{figure}
\centering
\includegraphics[scale=0.55]{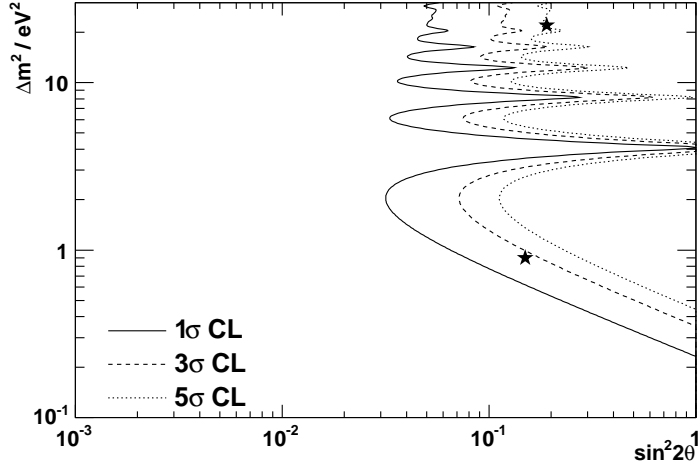}
\parbox{6in}{
\caption{\small Active-Sterile neutrino oscillation sensitivity with the NC reaction \nuxcarbon\
for SNS with a near detector, 5\% \fc\ systematic errors, and three years of running.  The two black stars correspond to the (3+2)
solutions (see text).}  
\label{f7}}
\end{figure}

\begin{figure}
\centering
\includegraphics[scale=0.55]{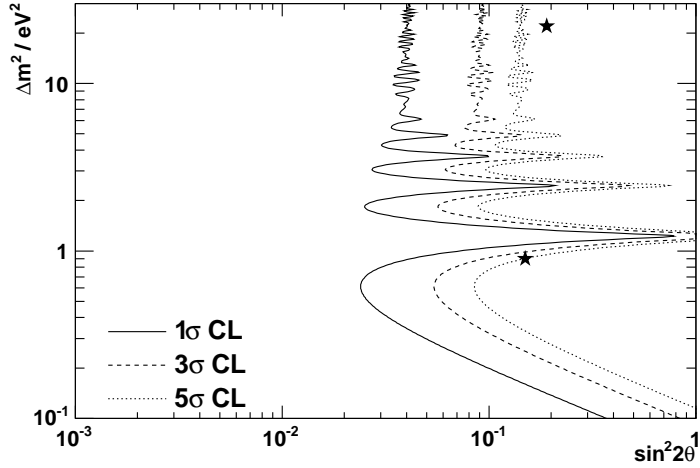}
\parbox{6in}{
\caption{\small Active-Sterile neutrino oscillation sensitivity with the NC reaction \nuxcarbon\
for SNS with a far detector, 5\% \fc\ systematic errors, and three years of running.  The two black stars correspond to the (3+2)
solutions (see text).}  
\label{f8}}
\end{figure}

If we consider just a single detector setup at the SNS, either near or
far, Figures \ref{f7} and \ref{f8} show what might be expected for 5\%
\fc\ errors.  Figure \ref{f7} shows that even this one near detector
has good sensitivity to active-sterile
oscillations with parameters of $\dm > 1$\,\ev\ and $\st > 0.10$ at $3
\sigma$.  This interesting measurement can be performed in parallel to
the neutrino carbon cross section measurement planned for the
$\nu$-SNS detector \cite{ornl}.

In figures 4 through 9, two black stars are plotted that correspond to
the (3+2) model sterile neutrino solution $\Delta \mbox{m} ^2 _{41}= 0.9$
\ev, $\sin^2 2 \theta_{41} = 0.15$, $\Delta \mbox{m} ^2 _{51}= 22$ \ev,
$\sin^2 2 \theta_{51} = 0.19$ \cite{(3+2)}.  This (3+2) model gives
significantly better fits than (3+1) models \cite{(3+2)}. Although the
sensitivity curves are strictly for (3+1) models and cannot be
directly applied to the (3+2) solutions, it can be seen that all
beam/detector combinations will be sensitive to these solutions. It
should be noted that the favored (3+1) solution from \cite{(3+2)} is
very close to the lower \dm\ (3+2) solution, and so this solution will
have coverage at the 3$\sigma$ to 5$\sigma$ level.  A recent (3+1)
analysis of all the short baseline data, excluding LSND, gives \mus\
oscillation plots that exclude the upper right regions shown here for
$\dm > 1 \ev$ and $\st > 0.1$ at the 99\% level
\cite{cirelli}.  Hence, even ignoring LSND, there is
much untested phase space that can be probed for active-sterile
neutrino oscillations.

\begin{figure}
\centering
\includegraphics[scale=0.55]{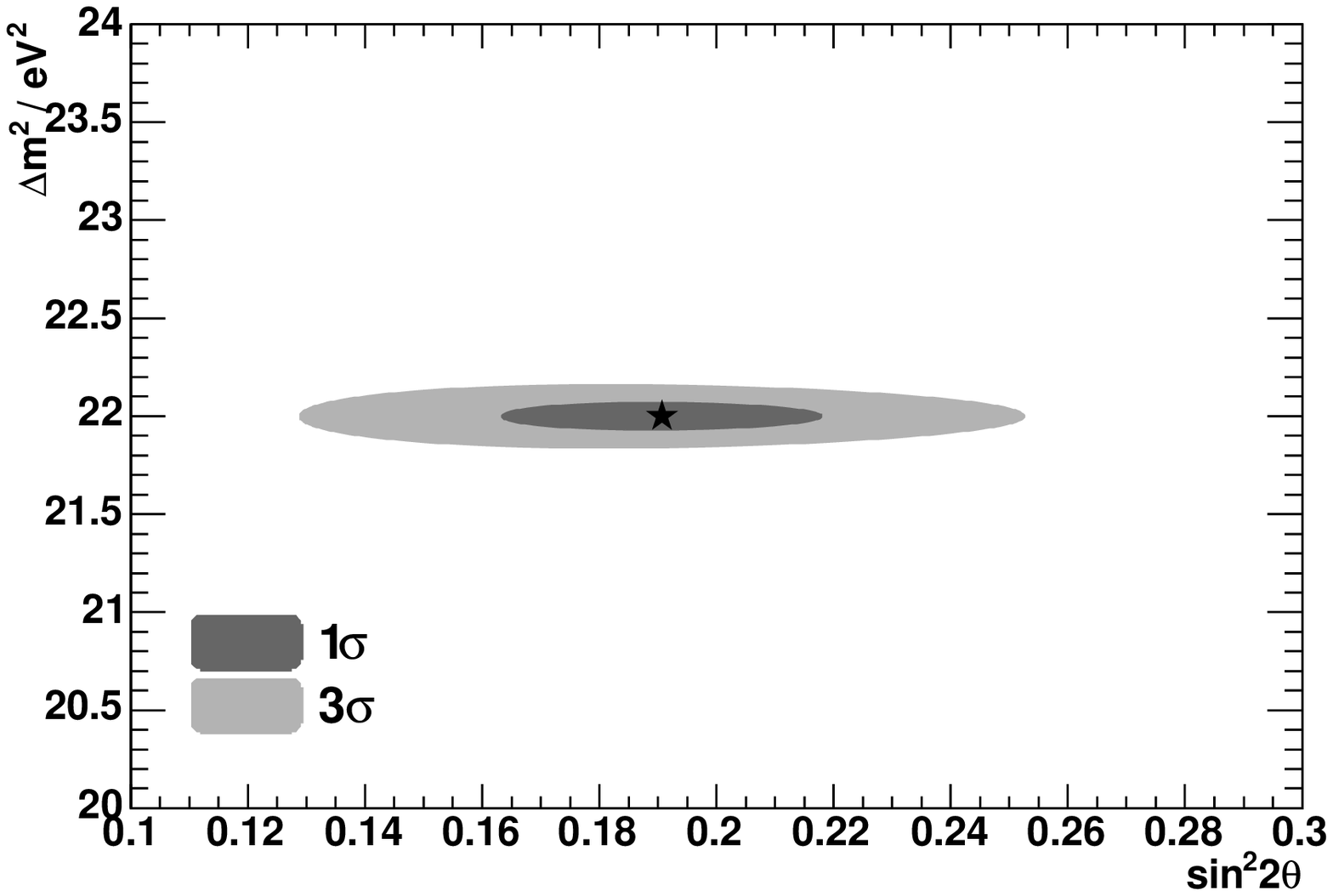}
\includegraphics[scale=0.55]{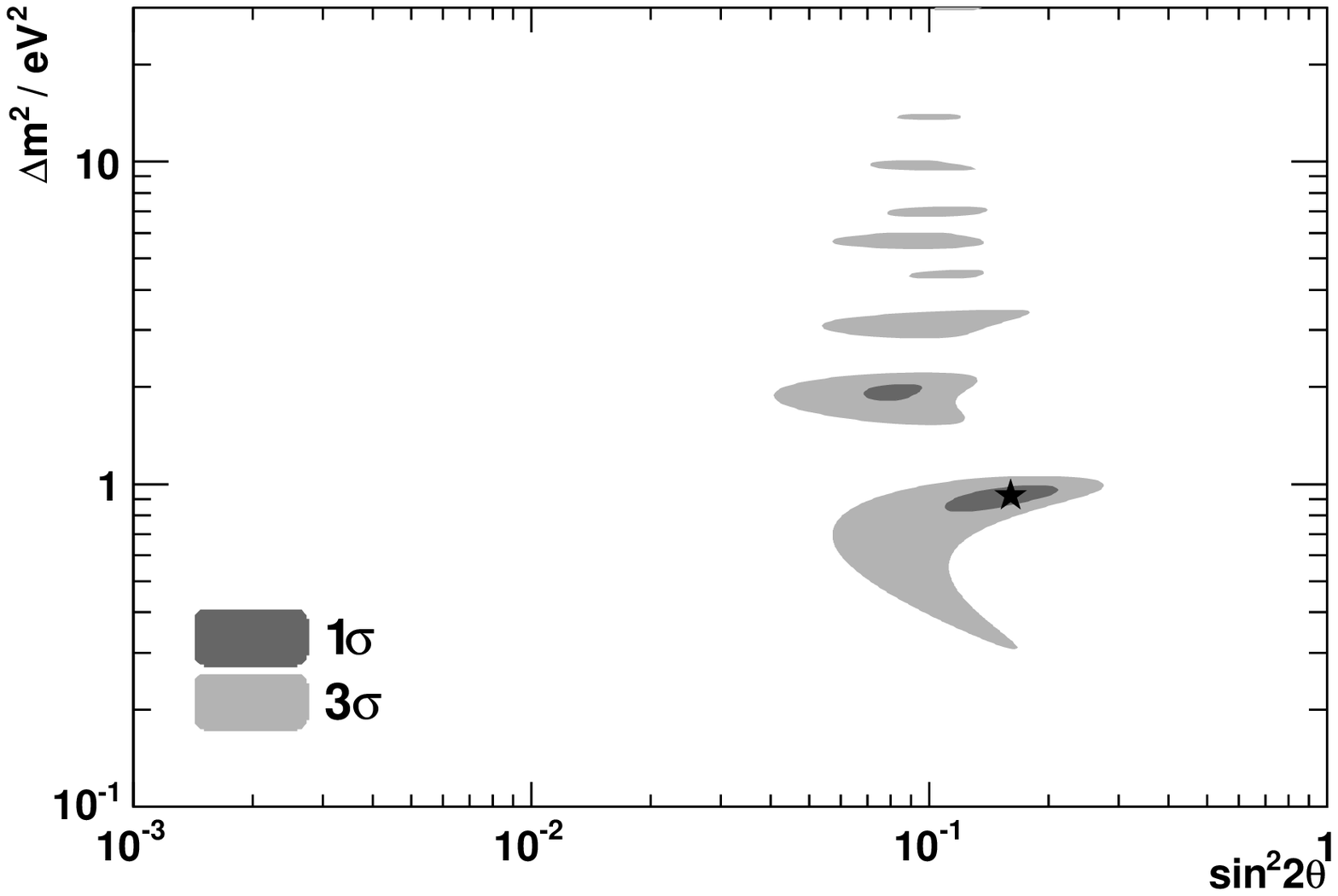}
\parbox{6in}{
\caption{\small The (3+2) active-sterile neutrino oscillation solution 
contours for SNS with two detectors, 5\% \fc\ systematic errors, 
and three years of running. See text for definition of $\st$.}
\label{f9}}
\end{figure}

\begin{figure}
\centering
\includegraphics[scale=0.55]{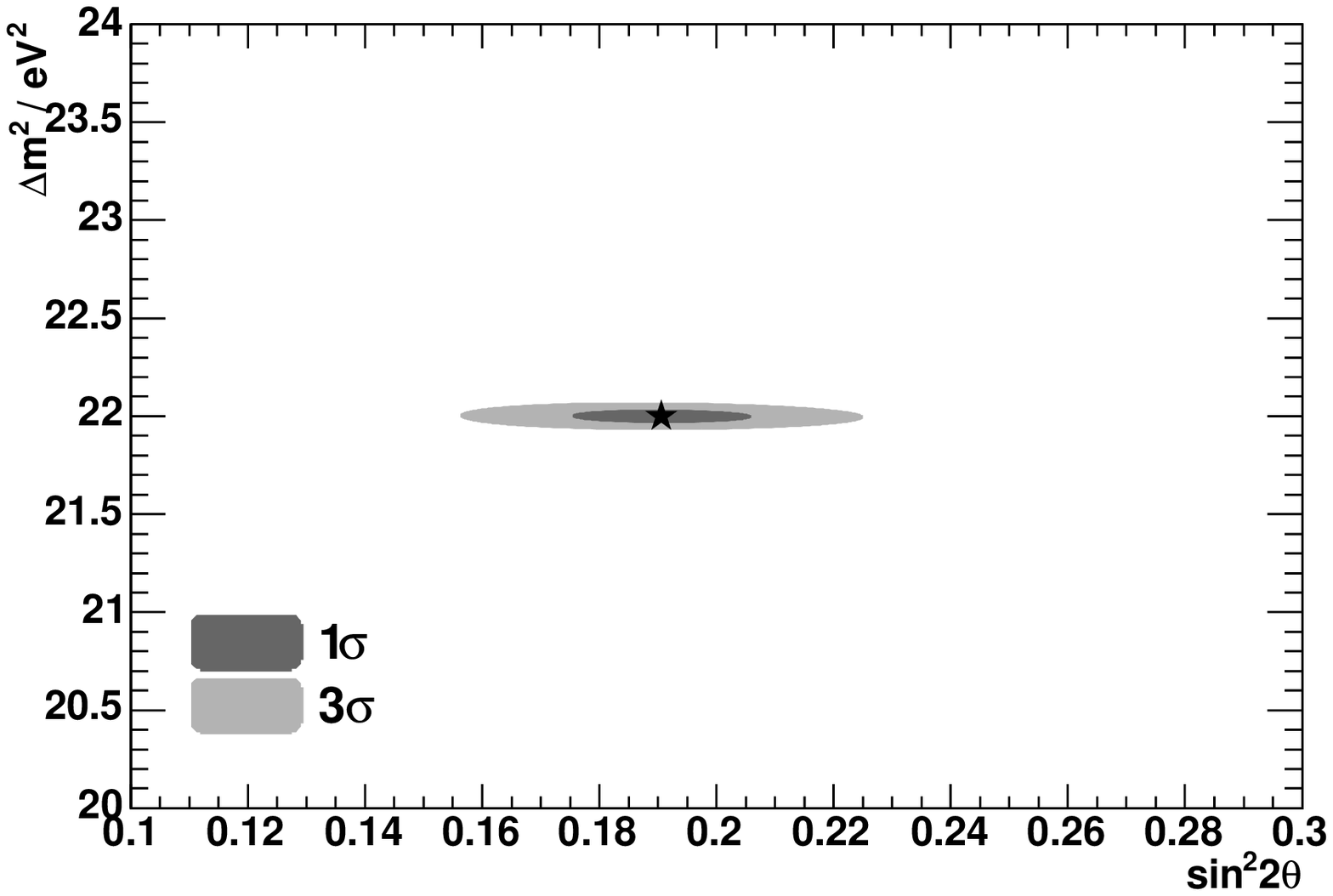}
\includegraphics[scale=0.55]{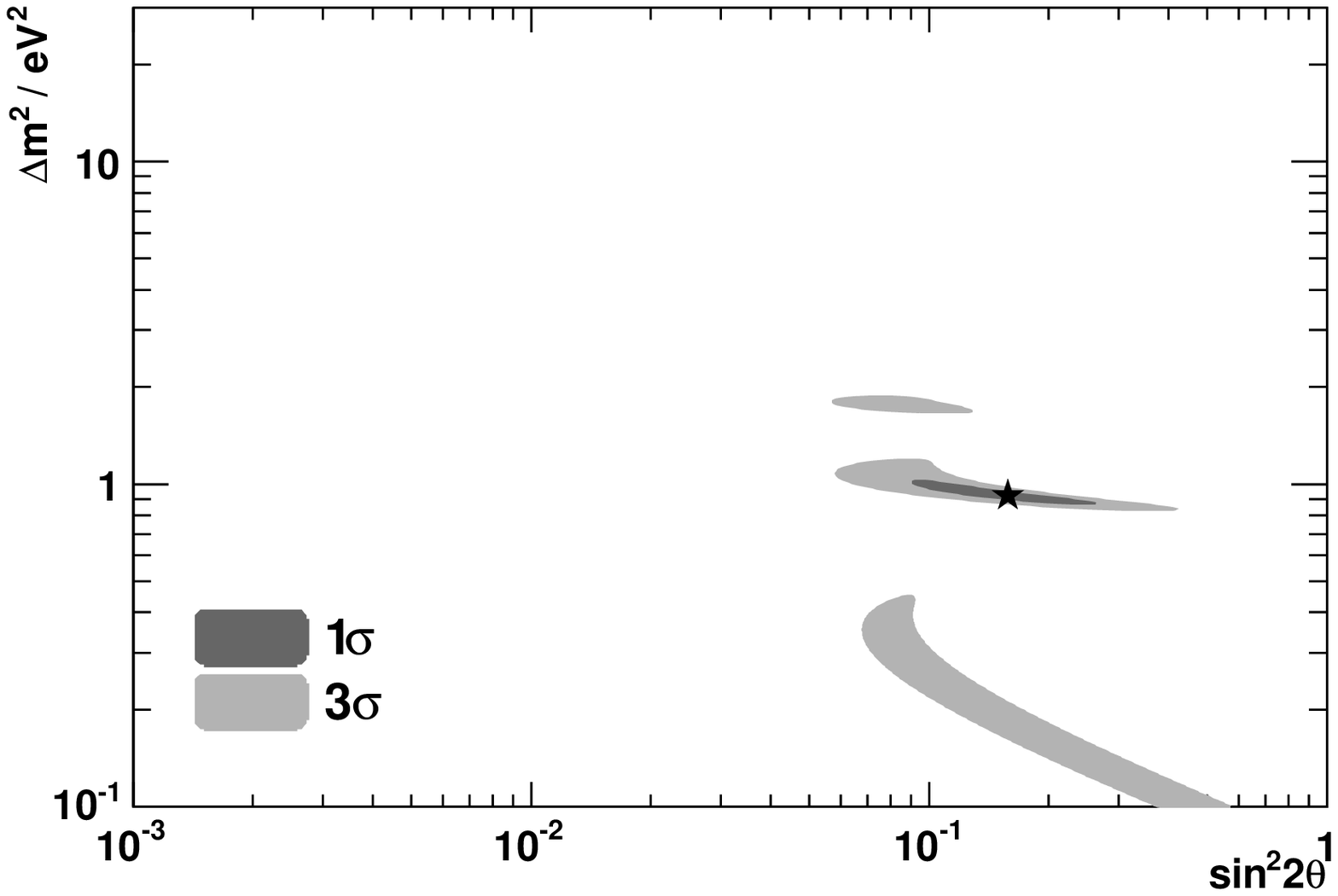}
\parbox{6in}{
\caption{\small The (3+2) active-sterile neutrino oscillation solution contours 
for FNAL with two detectors, 5\% \fc\ systematic errors, 
and three years of running.  See text for definition of $\st$.}
\label{f10}}
\end{figure}

Explicit treatment of (3+2) models is shown in figures \ref{f9} and
\ref{f10}, where we present parameter determination of the (3+2)
solutions. These models cannot be approximated by a two neutrino
approximation, and consequently $\st$ can no longer be interpreted as
before. We instead define $\st$ as $4\mbox{U}_{\mu n}^{2}(1-\mbox{U}_{\mu n}^{2})$,
where $\mbox{U}_{\mu n}$ is the relevant element in the neutrino mixing
matrix and $n$ is the sterile generation label. In these figures, one
of the two solutions is assumed and fixed.  This is a reasonable
procedure since the two solutions have quite different characteristics, i.e.\
the low \dm\ solution is demonstrated by a rate suppression 
between two detectors, while the high
\dm\ solution is demonstrated by rapid wiggles in L.  This allows for easy
separation of the two solutions.  The resulting plot, therefore, shows
the parameter determination at a single point in the full four
dimensional parameter space, rather than a projection in this
space. Both figures show that the parameter determination is
excellent.

\section{Conclusion}

In the coming year, new physics beyond the Standard Model will be
implied if MiniBooNE confirms the LSND oscillation anomaly.  To
understand whether active-sterile neutrino oscillations are
responsible for this new physics will require a new class of
oscillation experiments.  A stopped pion source such as the SNS,
coupled with two well positioned large liquid scintillator detectors,
can achieve the desired sensitivity.  This is made possible by the
fortuitous combination of a monoenergetic \numu\ flux and the efficiently
reconstructed neutral current interaction \nuxcarbon.  The SNS
two detector setup with 5\% \fc\ systematic errors can explore the \mus\ oscillation region 
$\dm > 0.4$\,\ev\ and $\st > 0.05$ at the 3$\sigma$ level.  
This would cover the \mus\
oscillation solutions indicated by LSND and the (3+2) model.  Such a
measurement would be the first clear indication of the existence of
light sterile neutrinos, and would herald a new era in neutrino
physics.

\section*{Acknowledgements}

We thank Yuri Efremenko, Tony Gabriel, and Michel Sorel for
helpful discussions.   This research was supported by LDRD
funding at LANL.

\newpage

\begin{thebibliography}{99}

\bibitem{mass} Y.~Fukuda {\it et al.}  [Super-Kamiokande Collaboration],
Phys.\ Rev.\ Lett.\  {\bf 81}, 1562 (1998) [arXiv:hep-ex/9807003];
Q.~R.~Ahmad {\it et al.}  [SNO Collaboration],
Phys.\ Rev.\ Lett.\  {\bf 89}, 011301 (2002) [arXiv:nucl-ex/0204008];
K. Eguchi {\it et al.} [KamlAND Collaboration], Phys.\ Rev.\ Lett.\ {\bf 90},
021802 (2003) [arXiv:hep-ex/0212021].


\bibitem{Gmann}M.\ Gell-Mann, P.\ Ramond, R.\ Slansky, in {\it SuperGravity}, ed.
by P.\ Van Nieuwenhuizen, D.Z.\ Freedman (North Holland, Amsterdam 1979), p.\ 315. But see, for example, 
G.\ J.\ Stephenson, Jr.\, T.\ Goldman, B.\ H.\ J.\  McKellar and M.\ Garbutt, 
hep-ph/0404015;  A.~de Gouvea, hep-ph/0501039.


\bibitem{noseesaw} A.Yu. Smirnov, arXiv:hep-ph/0411194.

\bibitem{(3+2)} M.\ Sorel, J.M.\ Conrad, M.H.\ Shaevitz, arXiv:hep-ph/0305255.

\bibitem{solar} P.C.\ de Holanda, A.Yu.\ Smirnov, arXiv:hep-ph/0307266.

\bibitem{supernova} A.B.\ Balantekin, G.M.\ Fuller, arXiv:astro-ph/0309519;
G.~C.~McLaughlin, J.~M.~Fetter, A.~B.~Balantekin and G.~M.~Fuller,
Phys.\ Rev.\ C {\bf 59}, 2873 (1999)
[arXiv:astro-ph/9902106].

\bibitem{pulsar} G.M.\ Fuller, A.\ Kusenko, I.\ Mocioiu, S.\ Pascoli, arXiv:astro-ph/0307267.

\bibitem{dmatter} 
K.~Abazajian, G.~M.~Fuller and M.~Patel,
Phys.\ Rev.\ D {\bf 64}, 023501 (2001)
[arXiv:astro-ph/0101524].


\bibitem{denergy} D.B.\ Kaplan, A.E.\ Nelson, N.\ Weiner, arXiv:hep-ph/0401099.
R.~Fardon, A.~E.~Nelson and N.~Weiner,
JCAP {\bf 0410}, 005 (2004); 
P.\ Q.\ Hung, hep-ph/0010126; P.\ Q.\ Hung, H.\ Pas, astro-ph/0311131;
G.\ J.\ Stephenson, 
Jr.\, T.\ Goldman and B.\ H.\ J.\ McKellar, Int.\ J.\ Mod.\ Phys.\ A13, 2765 (1998) [hep-ph/9603392]; 
  P.~Gu, X.~Wang and X.~Zhang,
  Phys.\ Rev.\ D {\bf 68}, 087301 (2003)
  [arXiv:hep-ph/0307148].


\bibitem{extraD}
  H.~Pas, S.~Pakvasa and T.~J.~Weiler,
  arXiv:hep-ph/0504096.


\bibitem{lsnd} A.~Aguilar {\it et al.}  [LSND Collaboration],
Phys.\ Rev.\ D {\bf 64}, 112007 (2001) [arXiv:hep-ex/0104049].

\bibitem{reactor}  Felix Boehm, nucl-ex/9906010.

\bibitem{miniboone} 
A.~O.~Bazarko  [BooNe Collaboration],
arXiv:hep-ex/9906003.

\bibitem{VanDalen} G.J.\ VanDalen, nucl-ex/0309014.

\bibitem{cpt} G.\ Barenboim, L.\ Borissov, J.\ Lykken, hep-ph/0212116;
V.\ Barger, M.\ Sorel, and K.\ Whisnant, in preparation.

\bibitem{lorentz} V.\ A.\ Kostelecky, M.\ Mewes, arXiv:hep-ph/0406255.

\bibitem{ornl} The Spallation Neutron Source (SNS) is an accelerator-based source being built in Oak Ridge, Tennessee, by the U.S.\ DOE, http://sns.gov/.
Also see http://www.phy.ornl.gov/workshops/sns2/ for details on the neutrino source and cross section detector $\nu$-SNS.

\bibitem{fnal} S.J.\ Brice, S.\ Geer, K.\ Paul, R.\ Tayloe, 
arXiv:hep-ex/0408135.

\bibitem{xsecarbon} M.~Fukugita, Y.~Kohyama and K.~Kubodera,
Phys.\ Lett.\ B {\bf 212}, 139 (1988).

\bibitem{carbondecay} G.\ Baym, {\it et al.} Nuclear Physics (NUPABL), Volume {\bf A433} (1985), No.\ 1, pg 61.



\bibitem{karmen} B. Armbruster [KARMEN Collaboration], Phys.\ Lett.\ {\bf B 423}
(1998), 15-20.

\bibitem{cirelli}
  M.~Cirelli, G.~Marandella, A.~Strumia and F.~Vissani,
  Nucl.\ Phys.\ B {\bf 708}, 215 (2005)
  [arXiv:hep-ph/0403158].


\end{thebibliography}
\end{document}